\documentclass[aps,twocolumn,showpacs]{revtex4}
\usepackage{graphics}
\usepackage{latexsym}

\usepackage{graphicx,epsfig}
\usepackage{psfrag}

\newcommand{\be}{\begin{equation}}
\newcommand{\ee}{\end{equation}}
\def\bq{\begin{eqnarray}}
\def\eq{\end{eqnarray}}

\begin{document}
\bibliographystyle{apsrev}
\def\half{{1\over 2}}
\def \D {\mbox{D}}
\def\curl {\mbox{curl}\,}
\def \ep {\varepsilon}
\def \lleq {\lower0.9ex\hbox{ $\buildrel < \over \sim$} ~}
\def \ggeq {\lower0.9ex\hbox{ $\buildrel > \over \sim$} ~}
\def\beq{\begin{equation}}
\def\eeq{\end{equation}}
\def\ber{\begin{eqnarray}}
\def\eer{\end{eqnarray}}
\def \apl {ApJ, }
\def \aps {ApJS, }
\def \pd {Phys. Rev. D, }
\def \prl {Phys. Rev. Lett., }
\def \pl {Phys. Lett., }
\def \np {Nucl. Phys., }
\def \l {\Lambda}

\title   {Cosmological Dynamics of Phantom Field}
\author{ Parampreet Singh }
\email{param@iucaa.ernet.in}
\author{M.Sami}
\altaffiliation[On leave from:]{ Department of Physics, Jamia Millia, New Delhi-110025}
\email{sami@iucaa.ernet.in}
\author{Naresh Dadhich}
\email[]{nkd@iucaa.ernet.in }
\affiliation{IUCAA, Post Bag 4, Ganeshkhind,\\
 Pune 411 007, India.}

\begin{abstract}
We study the general features of the dynamics of the phantom field in the cosmological context. In the case of inverse coshyperbolic potential,
we demonstrate that the phantom field can successfully drive the observed current accelerated expansion of the universe with the equation 
of state parameter $w_{\phi} < -1$.
The de-Sitter universe turns out to be the late time attractor of the model. The main features of the dynamics are independent of the initial
conditions and the parameters of the model. The
model fits the supernova data very well, allowing for $-2.4 < w_{\phi} < -1$ at
95 \% confidence level. 
\end{abstract}
 \maketitle
\section{introduction}

The observational evidence for accelerating universe has been for the past few years one of the central themes of  modern cosmology. The explanation of these observations in the framework of the standard cosmology requires an exotic form of energy which violates the
strong energy condition. A variety of scalar field models have been conjectured for this purpose including quintessence \cite{phiindustry}, K-essence \cite{Kes} and
recently tachyonic scalar fields\cite{ashoke,ashokeG,ashokeP,gibbons,tachyonindustry,tachyon1}. A different approach to the cosmic acceleration is advocated in Refs \cite{babak}.(For a review of the issues related to the cosmological constant and the dark energy, see \cite{ph2,sccc,ratra,tpcc}). By choosing an appropriate potential $V(\phi)$ and tuning the parameters of the model one can
account for the current acceleration of the universe with $\Omega_{\phi}=0.7$ and $\Omega_m=0.3$. However, all these models lead to the
equation of state parameter $w_{\phi} \geq -1$. The recent observations seem to favor the values of this parameter less than -1. \\

A scalar field with negative kinetic energy called the phantom field is proposed by Caldwell, Carroll et al and others to realize the possibility of late time acceleration with $w_{\phi} <-1$
 \cite{ph1,ph3,ph4,ph5,ph6,ph7,ph8,shatanov,ph9,ph10,hann,t1,t2,model,ph12,ph13,ph14,ph15,ph16,ph17,ph18,ph19,ph20,ph21,ph22}.
Such a field has a very unusual dynamics as it violates null dominant energy condition (NDEC). The models with equation of state parameter less than -1 are
known to face the problem of future curvature singularity which can however be overcome in specific models of phantom field. Inspite of
the fact that the field theory of phantom fields does encounter the problem of stability which one could try to bypass by assuming them to be effective fields\cite{t2,ph15,ph17}, it is nevertheless interesting to study their cosmological implications. Curiously, phantom fields have glorious
lineage in Hoyle's version of the Steady State Theory. In adherence to the Perfect Cosmological Principle, a creation field (C-field) was for the first time introduced \cite{h} to reconcile with homogeneous density by creation of new matter in the voids caused by the expansion of the universe. It was further refined and reformulated in the Hoyle and Narlikar theory of gravitation \cite{hn}, see 
Refs\cite{pn,nb,vishwa,vishwa1} for details of the C-field cosmology. The C-field appeared on the right hand side of the Einstein equation. What was conserved was the sum of the stress-tensors of matter and C-field and neither was  conserved separately. The C-field violated the weak energy condition. In this context, it may also be recalled that wormholes do require violation of energy conditions, in particular of the average null energy condition \cite{matt}. Interestingly, it has recently been shown that a very finite amount of exotic
energy condition violating matter is sufficient to produce a wormhole \cite{matt2}. Thus phantom fields though very exotic are not entirely new and out of place. Here the motivation is strongly driven by the observation.

\section{Dynamics of the Phantom Field}
The Lagrangian of the phantom field minimally coupled to gravity and matter sources is given by\cite{ph15}
\begin{equation}
{\cal{L}} =(16 \pi G)^{-1} R+{1 \over 2} g^{\mu \nu} \partial_{\mu} \phi \partial_{\nu} \phi - V(\phi)+{\cal{L}}_{source}
\label{lagrang}
\end{equation}
where ${\cal{L}}_{source}$ is the remaining source term (matter,radiation) and $V(\phi)$ is the phantom potential. The kinetic energy term 
of the phantom field in (\ref{lagrang})  enters with the opposite sign in contrast to the ordinary scalar field (we employ the
metric signature, -,+,+,+). It is the negative kinetic energy that distinguishes phantom fields from the ordinary fields. The Einstein 
equations which follow from (\ref{lagrang}) are
\begin{equation}
R_{\mu \nu}-{1 \over 2}g_{\mu \nu}R=T_{\mu \nu}
\label{einsteineq}
\end{equation}
with
\begin{equation}
T_{\mu \nu}=T_{\mu \nu}^{source}+T_{\mu \nu}^{ph}
\label{stress}
\end{equation} 
where $T^{ph}_{\mu \nu}$ and $T_{\mu \nu}^{source}$ are the stress tensors of phantom field and of the background (matter,radiation).
In contrast to the case of the creation field for which the individual stress-tensors could not be
conserved independently, the phantom field energy momentum tensor $T_{\mu \nu}^{ph}$ here and the usual matter field stress-tensor $T_{\mu \nu}^{source}$
are indeed conserved separately. The former was introduced to create new
matter out of nothing so as to keep the density uniform in an expanding
universe. The latter has no such purpose instead its job is to accelerate
expansion at late times with $w_{\phi}<-1$.

The stress tensor for the phantom field which follows from
(\ref{lagrang}) has the form

\begin{equation}
T_{\mu \nu}^{ph}=- \partial_\mu \phi \, \partial_\nu \phi + g_{\mu \nu}\left[{1 \over 2}g^{\alpha \beta} \, \partial_\alpha \phi \, \partial_\beta \phi
-V(\phi) \right]
\label{stress1}
\end{equation}
We shall assume that the phantom field is evolving in an isotropic and homogeneous space-time and that $\phi$ is a function of
time alone. The energy density $\rho_{\phi}$ and pressure $p_{\phi}$ obtained from $T_{\mu \nu}^{ph}$ are 
\begin{equation}
\rho_{\phi}=-{\dot{\phi}^2 \over 2}+V(\phi),~~~~~p_{\phi}=-{\dot{\phi}^2 \over 2}-V(\phi).
\label{ke}
\end{equation}
The fact that the phantom field feels, in contrast to ordinary field/matter, opposite curvature of the space-time is reflected in the negativity of kinetic energy in eqn. (\ref{ke}). The equation of state parameter is now given by
\begin{equation} 
w_{\phi}={{{\dot{\phi}^2 \over 2}+V(\phi)} \over { {\dot{\phi}^2 \over 2}-V(\phi) }}.
\end{equation} 
Now the conditions $w_{\phi} <-1, ~~ \rho_{\phi} >0$ would prescribe the range,
$0 < \dot{\phi}^2 < 2V(\phi)$.
The Friedman equation which follows from eqn. (\ref{einsteineq}) with the modified energy momentum tensor given by eqn. (\ref{stress}) is
\begin{equation}
{\dot{a}^2(t) \over a^2(t)}={8 \pi G \over 3}\left[\rho_{\phi}+\rho_b \right]
\label{friedman}
\end{equation}
where the background energy density due to matter and radiation is given by
\begin{equation}
\rho_b={\rho^i_R \over a^4}+{\rho^i_m \over a^3}
\end{equation}
The peculiar nature of the phantom field also gets reflected in its evolution equation which follows from eqn. (\ref{lagrang})
\begin{equation}
\ddot{\phi}+{3\dot{a} \over a}\dot{\phi}=V_{,\phi}(\phi).
\label{evoleq}
\end{equation}
Note that the  evolution equation (\ref{evoleq}) for the phantom field is same as that of the normal scalar field with the inverted potential 
allowing the field with zero initial kinetic energy
to roll up the hill; i.e. from lower value of the potential to higher one. At the first look, such a situation seems to be pathological.
However, at present, the situation in cosmology is remarkably tolerant to any pathology if it can lead to a viable cosmological model.\par
The conservation equation formally equivalent to eqn. (\ref{evoleq}), has the usual form
\begin{equation}
\dot{\rho_{\phi}}+3H(\rho_{\phi}+p_{\phi})=0
\label{conseqn}
\end{equation}
and the evolution of energy density is given by
\begin{equation}
\rho_{\phi}=\rho_{0\phi}e^{-\int{6\left (1-\zeta(a)\right){da \over a}}}
\end{equation}
with
$$ \zeta(a)={1 \over {(K_e/ P_e) +1}}$$
where the ratio of  kinetic to potential energy$ (K_e/P_e)$ is given by
\begin{equation}
{K_e \over P_e}=-{\dot{\phi}^2 \over {2 V(\phi)}}.
\end{equation}
The evolution of potential to kinetic energy ratio plays a significant role in  the growth or decay of the energy density $\rho_\phi$
at a given epoch and will be crucial in the following discussion \cite{sami}.

\begin{figure}
\resizebox{3.0in}{!}{\includegraphics{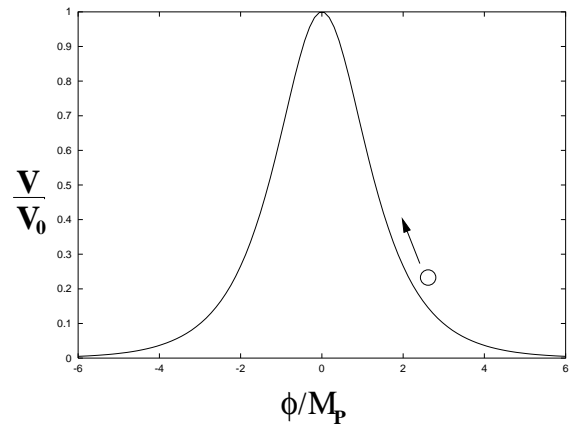}}
\caption{Evolution of the phantom field is shown for the model described by eqn. (\ref{pot}). Due to the unusual behavior, the phantom
field, released with zero kinetic energy away from the origin, moves towards the top of the potential. It sets into the damped oscillations about $\phi=0$ and ultimately settles there permanently.
}
\label{figpot}
\end{figure}

 Let us next address the question of choice of the potential of the scalar field which would lead to a viable cosmological
 model with $w_{\phi}<-1$. An obvious restriction on the evolution  is that
   the scalar field should survive till
today (to account for the observed late time accelerated expansion)
 without interfering with the nucleosynthesis of the standard model and on the other hand should also avoid the future collapse of the universe. It was indicated
by Carroll, Hoffman and Trodden \cite{t2} as how to build  models free from future singularity with $w_{\phi} < -1$.
In this paper we examine a model with the scalar phantom field
 which leads to a viable cosmology with $w_{\phi} < -1$ and shortly after driving the current acceleration of universe the field settles at $w_{\phi}=-1$ thereby avoiding the future singularity. We confront the model with supernova Ia
observations to constrain its parameters. The model fits the supernova
data quite well for a large range of parameters.

\section{A Viable Model with the Equation of State, $w_{\phi} < -1$}
\begin{figure}
\resizebox{3.0in}{!}{\includegraphics{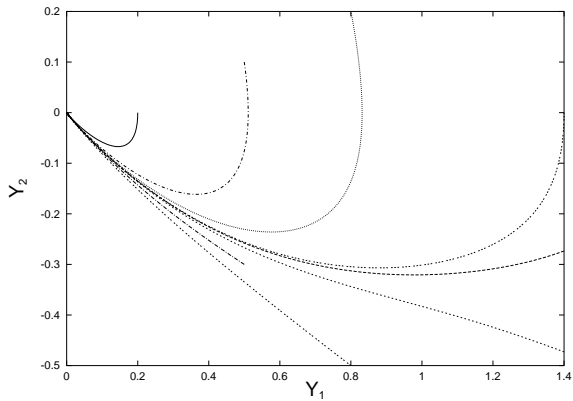}}
\caption{
Phase portrait (plot of $Y_2 \equiv \dot{\phi}/M_p^2$ versus $Y_1 \equiv \phi/M_p$) of the model described by eqn. (\ref{pot}). Trajectories
starting anywhere in the phase space end up at the stable critical point (0,0).}
\label{figphase}
\end{figure}

We shall here consider a model with
\begin{equation}
V(\phi)=V_0\left[\cosh \left({{\alpha \phi} \over M_p }\right)\right]^{-1}.
\label{pot}
\end{equation}
Due to its peculiar properties, the phantom field, released at a  distance from the origin with zero kinetic energy, moves
towards the top of the potential and crosses over to the other side and turns back to execute the damped oscillation about the
maximum of the potential (see figure \ref{figpot}). After a certain period of time the motion ceases and the field settles on the top of the potential
permanently to mimic the de-Sitter like behavior ($w_{\phi}=-1$). Indeed, the de-Sitter like phase is a late time
attractor of the model (see Fig. \ref{figphase}). It should be emphasized that these are the general features of phantom dynamics
which are valid for any bell shaped potential, say, $V(\phi)=V_0\left[1+\alpha \phi^2/M_p\right]^{-1}$ or the Gaussian potential
considered in Ref. \cite{t2}.

\begin{figure}
\resizebox{3.0in}{!}{\includegraphics{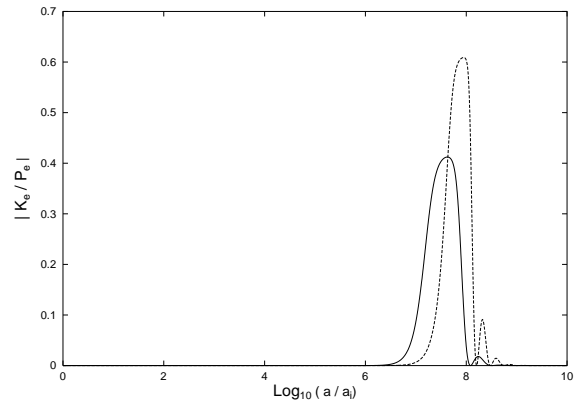}}
\caption{The ratio of kinetic to potential energy of the phantom field is plotted for the potential given by eqn. (\ref{pot}) for $\alpha=2$ (solid line) and $\alpha=3$ (dashed line). The evolution of $|K_e/P_e|$ starts later but peaks higher for larger value of $\alpha$. The
height of the peak is independent of $V_0$. The change in the value of $V_0$ merely shifts the position of the peak.}
\label{figKE}
\end{figure}
In order to investigate the dynamics described by eqns. (\ref{friedman}) and (\ref{evoleq}), it would be convenient to cast these equations as a system of first order equations
\begin{equation}
Y_1'=\frac{1}{H(Y_1,Y_2)} \, Y_2
\label{evol1d}
\end{equation}
\begin{equation}
Y'_2= -3Y_2+{1 \over H(Y_1, Y_2)}\Big[{d {\cal V}(Y_1) \over dY_1} \Big]
\label{evol2d}
\end{equation}
where
 \begin{equation}
 Y_1={\phi \over M_p},\quad Y_2={\dot{\phi} \over M_p^2},\quad {\cal V}={ V(Y_1) \over M_p^4}
 \end{equation}
and prime denotes the derivative with respect to the variable $N=\ln(a)$. The function
$ H(Y_1, Y_2)$ is given:
\begin{equation}
 H(Y_1, Y_2)=\sqrt{{1 \over 3}\left[{Y_2^2 \over 2}+ {\cal V}(Y_1) +{\rho_b \over M_p^4}  \right]} \label{hubble}
\end{equation}
where $\rho_b =\rho^i_r e^{-4N}+\rho^i_m e^{-3N}$. Using eqns. (\ref{pot}), (\ref{evol1d}) and  (\ref{evol2d}), it is not difficult to see that $(Y_1, Y_2)=(0, 0)$
is a fixed point of the system. Numerical analysis confirms the stability of the fixed point (Fig. \ref{figphase}). Thus the de-Sitter like
solution is the late time attractor of the model.\par
As for the initial conditions, for convenience we shall set them  in the radiation dominated era
with $a_{i}=1$ and $\rho^i_r=1~MeV^4$. Note that at the present epoch, the scale factor $a$ would  nearly be~  $4\times 10^{9}$.
The initial value for $\phi$ as well as the values of parameters in the potential ($\alpha$, $V_0$)
are chosen so as to ensure a viable cosmological model. 
These choices could be construed as fine tuning in the model. 
But it is no worse than the
fine tuning in some models of quintessence, for instance, see Ref.\cite{sami}.
It should however be noted that quintessence models based on tracker potentials
are independent of initial conditions and only require fine tuning of parameters of the potential.

\begin{figure}
\resizebox{3.0in}{!}{\includegraphics{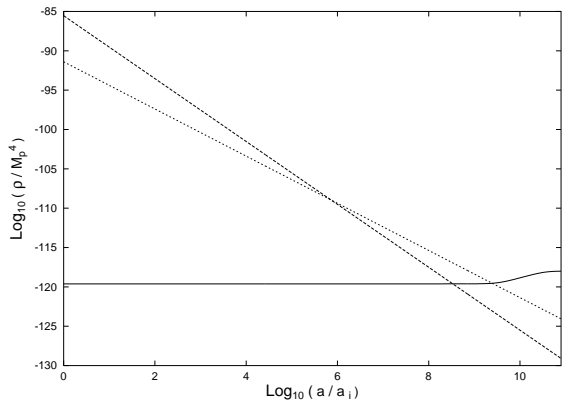}}
\caption{ The energy density is plotted against the scale factor: solid line corresponds to $\rho_{\phi}$ for $\alpha=1.26$ in case
of the model (\ref{pot}) with $V_0^{1/4}\simeq 3\times 10^{-30} M_p $. The dashed and dotted lines  correspond to energy density of radiation and matter.
Initially, the energy density of the phantom field is extremely subdominant and remains to be so for most of the period of evolution.
At late times, the field energy density catches up with the background, overtakes it and starts growing ($w_{\phi)}<-1)$ and drives
the current accelerated expansion of the universe before freezing to a constant value equal to $-1$ (in future).}
\label{figden}
\end{figure}

\begin{figure}
\resizebox{3.0in}{!}{\includegraphics{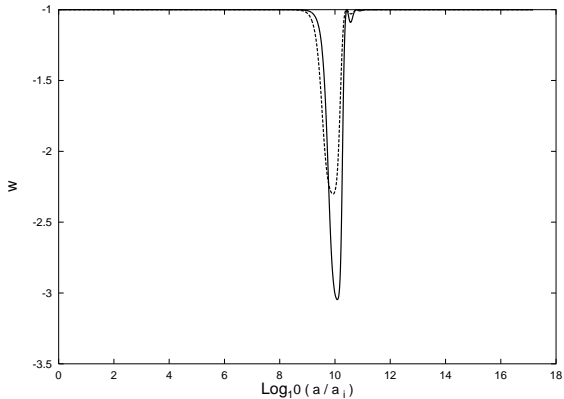}}
\caption{Evolution of the equation of state parameter $w_{\phi}$ is shown as a function of the scale factor for the model
described in figure \ref{figden} 
in case of $\alpha=2.5$ (solid line) and $\alpha=2$ (dashed line) . Except for a short period, $w_{\phi}$ is seen to be constant (-1). The parameter evolves to negative values less than -1 (smaller for larger value of $\alpha$) at late times leading to current acceleration of the universe . It then executes damped oscillations and fast stabilizes to $w_{\phi}=-1$.}
\label{figstate}
\end{figure}
 
\begin{figure}
\resizebox{3.0in}{!}{\includegraphics{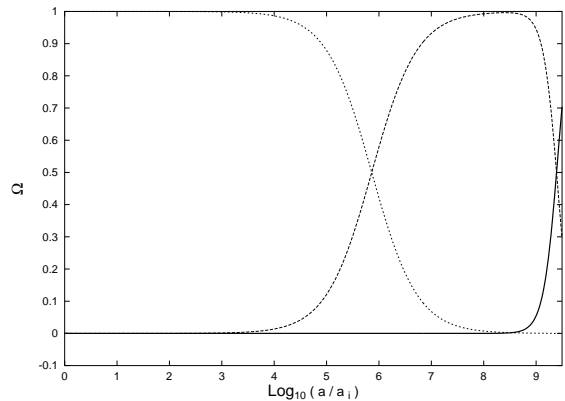}}
\caption{Dimensionless density parameter $\Omega$ is plotted against the scale factor for the model described by eqn. (\ref{pot})  with $\alpha=1.26$ and $ V_0^{1/4} \simeq 3\times 10^{-30}M_p$ for: (i) phantom field (solid line), (ii) radiation (dotted line) and matter (dashed line). Late time behavior of the phantom field
leads to the present day value of $\Omega_{\phi}=0.7$ and $\Omega_m=0.3$ for the equation of state parameter $w_{\phi}<-1$.}
\label{figOmega}
\end{figure}
We shall now describe the dynamics of the model due to the potential (\ref{pot}). Initially, the field is displaced from the maximum of the potential
and its energy density is subdominant with respect to the background energy density. As a result the background plays the deciding role in the evolution
dynamics. The Hubble damping due to $\rho_b>>\rho_{\phi}$ is (extremely) large in the field evolution equation. Consequently, the
field does not evolve and freezes at the initial position mimicking the cosmological constant like behavior. Meanwhile, the background
energy density red shifts as $1/a^n$ (n=4 for radiation). The phantom field $\phi$ continues in the state with $w_{\phi}=-1$ till the moment
$\rho_{\phi}$ approaches $\rho_b$. The background ceases now to play the leading role (becomes subdominant) and the phantom field takes over and it moves fast towards the
top of the potential. The kinetic to potential energy ratio $|K_e/P_e|$ rapidly increases and becomes maximum at $\phi=0$ allowing 
the equation of state parameter  to attain the  minimum (negative) value. This leads to the fast increase in $\rho_{\phi}$. Damped oscillations then set in the system making the ratio to oscillate to zero (see Fig. \ref{figKE})  and allowing  $\rho_{\phi}$ to settle 
ultimately at a constant value ($w_{\phi}=-1$) for ever. This development is summarized in Figs. (\ref{figden}) and 
(\ref{figstate}) . As shown in Fig. (\ref{figKE}), the ratio $|K_e/P_e|$ takes off later but peaks higher for larger value of the parameter $\alpha$. The
reason is that for larger value of  $\alpha$ the field energy density is lower and consequently it takes longer for $\rho_{\phi}$
to catch up with the background. Once the background value is reached, the field sets into motion and rolls towards the maximum of the potential and the roll-up is faster as the potential gets steeper;i.e. larger is the value of $\alpha$. We should however emphasize that the main features of the evolution are absolutely independent
of the initial conditions and the values of the parameters in the model. However, tuning of the parameters is required to get the right things to happen at right time. For the major period of time the equation of state, $w_{\phi} = -1$ while for relatively short time
$w_{\phi}$ is required to be $<-1$ (see Fig. (\ref{figstate})). The later happens when $\rho_{\phi}$ overtakes the background (at late times) and starts growing, leading to the fast
growth of $\Omega_{\phi}$. By tuning the parameters of the model it is possible to account for the current accelerated expansion of the universe with $\Omega_{\phi}=0.7$ and $\Omega_m=0.3$ during the period when $w_{\phi}<-1$ ( see Fig. \ref{figOmega}).

\section{Constraints on parameter space from Supernova observations}

We use the Supernova Ia observations to put constrains on the parameter
space of the phantom field model. For that let us first note that the 
luminosity distance ($d_L$) for a source at redshift $z$ located at 
radial coordinate distance $r$ is given by $d_{\rm L} = (1 + z) \,a_0 \, r$,
where $a_0$ is the present value of the scale factor.
This can be used to define the dimensionless luminosity distance, 
 ${\cal D}_{\rm L} \equiv H_0 d_{\rm L}$, where $H_0$ is the present value of 
the Hubble parameter. The apparent magnitude ($m$) of the 
source is given as  
\be
m (z) = {\cal M} + 5 \, \log [{\cal D}_{\rm L}(z) ] \label{eq:mageq}
\ee
where ${\cal M} \equiv M - 5 \,\log H_0 +$ constant.

We use the magnitude-redshift data of 57 supernovae of type Ia, which include
54 supernovae considered by Perlmutter et al (excluding 6 outliers from the 
full sample of 60 supernovae) \cite{perl}, SN 1997ff at $z = 1.755$ 
\cite{riess} and two
newly discovered supernovae SN2002dc at $z = 0.475$ and SN2002dd at $z = 0.95$
\cite{blakeslee}. The best fitting values of the parameters can be obtained
through $\chi^2$ minimization where,
\be
\chi^2 = \sum_{i = 1}^{57} \, \Bigg[\frac{m_i^{\rm eff} - m(z_i)}{
\delta m_i^{\rm eff}} \Bigg]^2 ~.
\ee
Here $m_i^{\rm eff}$ refers to the effective
magnitude of the $i$th supernovae which has been corrected by the lightcurve
width-luminosity correction, galactic extinction and the K-correction.
The uncertainty in $m_i^{\rm eff}$ is denoted by $\delta m_i^{\rm eff}$.

\begin{figure}[tbh!]
\epsfig{figure=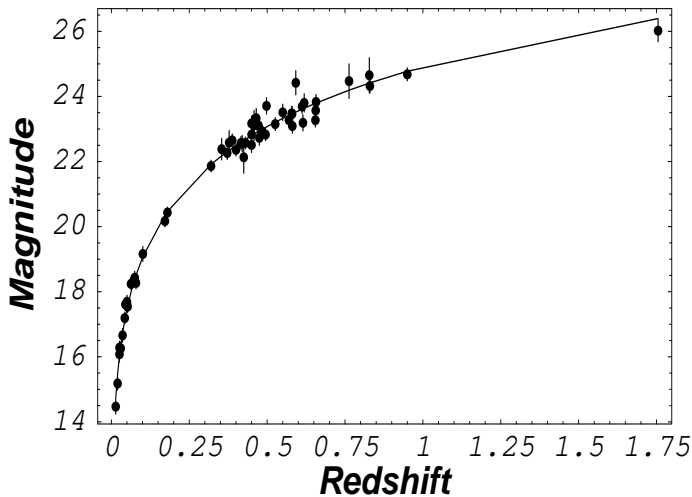,height=2.6in,width=3.6in,angle=0}
{\caption{\small The magnitude-redshift plot for 57 supernovae, including the
recently discovered, SN2002dc and SN2002dd. The theoretical curve is for the
best fit value obtained at $\alpha = 2.3$ and $\Omega_{V_0} \equiv 3 V_o/\rho_c = 1500$ for $\phi_i = 3.5 M_P$.   }}
\end{figure}

Since our aim is to constrain parameter space of the phantom field 
we tuned the $\rho_m$ at initial epoch in such a way that in the
absence of phantom dynamics it yields
the ratio of matter density to critical density ($\rho_c$) today as 
$\Omega_{m} = 0.3$ which is consistent with other observations
including $WMAP$ \cite{wmap}. However, since the equations for the
matter and radiation density evolution and the phantom dynamics are
coupled, not all values of the parameters $V_0$ and $\alpha$ would
yield $\Omega_m = 0.3$ and $\Omega_\phi = 0.7$ today. Apart from 
these two parameters, there are
initial conditions on $\phi$ and $\dot \phi$ of which the latter was fixed 
to zero (in fact we found that results are not affected on reasonable 
variation of this initial condition). 
\begin{figure}[tbh!]
\epsfig{figure=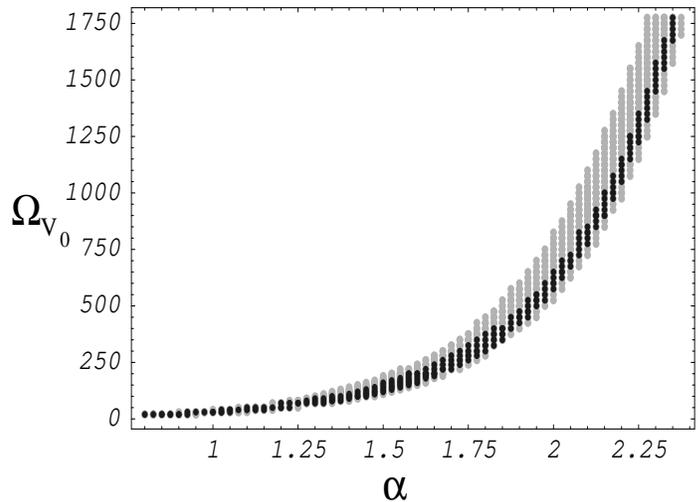,height=2.6in,width=3.6in,angle=0}
{\caption{\small The 68.3 \% and 99.7 \% confidence regions of the
phantom parameter space are shown as black and grey regions respectively}}
\end{figure}
For a viable evolution of the 
universe it is required that initial value of $\phi$ ($\phi_i$) be at least of
 the order of few. In case of $\phi_i$  much larger than unity, the equation of
state turns out to be close to -1 for admissible values of $\Omega_m$ and
$\Omega_{\phi}$, thus reasonable values of $\phi_i$ are of the 
order of few which
yield present value of the equation of state parameter less than -1.
The best fitting parameters after $\chi^2$ minimization were found to be 
$\alpha = 2.3$, 
${\Omega_{V_0}} = 1500$ ($\Omega_{V_0} \equiv 3 V_0/\rho_c$), $\phi_i = 3.5 M_P$ and  ${\cal M} = 23.8$  with $\chi^2 = 61.9$
and $\chi^2$ per degrees of freedom as 1.17 which represents a good fit, as
is shown in Fig. 7. 
 For the same data set the $\chi^2$ per degree of freedom for the best fitting flat model for the constant equation of state turns out to 
be 1.08 with $\Omega_m = 0.32$ \cite{vishwa2}.
The best fit phantom parameters yield an accelerating universe with 
 present values of the equation of state 
as   $w_\phi = - 1.74$, $\Omega_m = 0.3$ and $\Omega_\phi = 0.7$ 

In order to constrain the parameter space we fixed the arbitrariness in
$\phi_i$ to its best fit value and marginalized over ${\cal M}$. In Fig. 8
we have shown the 68.3 $\%$ and 99.7 $\%$ confidence regions in the 
parameter space of $\Omega_{V_0}$ and $\alpha$. As depicted in the figure,
a large region of the parameter space is allowed by the supernova observations. However, we should emphasize that  a small change in $\alpha$ 
corresponds to a large variation in $\Omega_{V_0}$ which reflects a fine
tuning of parameters  similar to other models of dark energy.
 Fig. 9
depicts  95.4 \% confidence region of the parameter space with different
allowed ranges of the equation of state. Dark energy models with the  
equation of state less than -1 have recently been analyzed and  the bounds
have been obtained \cite{hann,t1,t2,model,ph19}. At 95.4 \% confidence level the bound on
$w_\phi$ for the model under consideration is found to be -2.4 $ < w_\phi < $ -1. 
Thus, the phantom field model being constrained by supernova observations,
favors a lower value of equation of state parameter. 
Similar bounds have been obtained for other models with phantom energy \cite{hann}.

\begin{figure}[tbh!]
\epsfig{figure=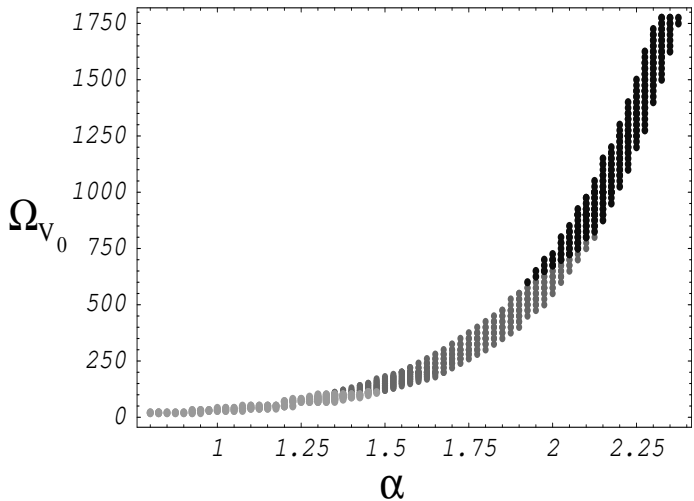,height=2.6in,width=3.6in,angle=0}
{\caption{\small The figure shows 95.4 \% confidence region of the phantom
parameter space. The light gray shaded region shows allowed parameters 
which yield $w_\phi > -1.3$, the dark shaded gray region corresponds to
$ -1.3 < w_\phi < -1.6$, whereas the black region represents $w_\phi < -1.6$.  }}
\end{figure}

\section{Discussion}

In this paper we have investigated the general features of the cosmological dynamics of the phantom field. In the case of the inverse coshyperbolic 
potential, we have shown that the phantom field can account for the current acceleration of universe with the negative values of
the equation of state parameter ($w_\phi < -1$). The general features of the model are shown to be independent of the
initial conditions and the values of the parameters in the model. However, tuning of the parameters is necessary for a viable evolution. 
Unlike, the quintessence models based on tracker potentials which need
fine tuning of only potential parameters, the phantom field model also requires
fine tuning of the initial condition of the field to account for the
current acceleration of the universe.
The model fits supernova
Ia observations fairly well for a certain range of parameters. The best fitting parameters of the model correspond to the equation of state parameter $w_\phi = -1.74$ and can vary up to -2.4 at 95.4 \% confidence level. In our analysis we have investigated a particular model which
avoids future singularity. It is expected that other models of this class (as mentioned above) would exhibit  similar behavior.
It would be interesting to  
further constrain the parameter space by other
observations like CMB and structure formation.

\begin{acknowledgements}
We thank T. R. Choudhury, Jian-gang Hao, B. Mcinnes, T. Padmanabhan and V. Sahni for useful discussions. we are also thankful to Sean Carroll for helpful comments
on the first draft of the paper.
PS thanks CSIR for a research grant.
\end{acknowledgements}


\begin{thebibliography}{99}
\bibitem{phiindustry}
B. Ratra, and P.J.E. Peebles, {\it Phys.~Rev.~D~} {\bf 37}, 3406 (1988);
C. Wetterich, Nucl. Phys. B {\bf 302}, 668 (1988);
J. Frieman, C.T. Hill, A. Stebbins, and I. Waga, (1995)
{\it Phys.~Rev. Lett.~} {\bf 75}, 2077;
P.G. Ferreira and M. Joyce, {\it Phys.~Rev.~D~} {\bf 58}, 023503 (1998);
I. Zlatev, L, Wang and P.J. Steinhardt {\it Phys.~Rev. Lett.~} {\bf 82},
896 (1999);
P. Brax and J. Martin {\it Phys.~Rev.~D~} {\bf 61}, 103502 (2000);
  L.A. Ure\~{n}a-L\'{o}pez and T. Matos, {\it Phys.~Rev.~D~}
{\bf 62}, 081302 (2000);
T. Barreiro, E.J. Copeland and N.J. Nunes {\it Phys.~Rev.~D~} {\bf 61},
127301 (2000);
A. Albrecht and C. Skordis {\it Phys.~Rev. Lett.~} {\bf 84}, 2076 (2000); V. Johri, astro-ph/0005608; V. Johri, astro-ph/0108247; V. Johri, astro-ph/0108244; 
J. P. Kneller and  L. E. Strigari, asto-ph/0302167; F. Rossati, hep-ph/0302159; V. Sahni, M. Sami and T. Souradeep, Phys.Rev. D65 (2002) 023518[gr-qc/0105121]; M. Sami, N. Dadhich and Tetsuya Shiromizu, hep-th/0304187.
\bibitem{Kes} Armendariz-Picon, T. Damour, V. Mukhanov, Phys.Lett. B458 (1999) 209 [hep-th/9904075];
T. Chiba, T. Okabe and M. Yamaguchi, Physical Review D{\bf 62}, 023511 (2000).
\bibitem{ashoke}
A. Sen, arXiv: hep-th/0203211; arXiv: hep-th/0203265; arXiv: hep-th/0204143
and references cited therein.
\bibitem{ashokeG} M. R. Garousi, Nucl. Phys. B{\bf 584}, 284(2000); M. R. Garousi, hep-th/0303239.
\bibitem{ashokeP}E.A. Bergshoeff, M. de Roo, T.C. de Wit, E. Eyras, S. Panda, JHEP 0005 (2000) 009. 
\bibitem{gibbons}Gibbons, G.W, arXiv:hep-th/0204008.
\bibitem{tachyonindustry} 
 Fairbairn M and M.~H.~Tytgat, arXiv:hep-th/0204070; Feinstein, A., hep-th/0204140.
 Mukohyama, S arXiv:hep-th/0204084;
 Frolov, A, L.~Kofman and A.~Starobinsky, Phys.Lett. B{\bf 545}, 8 (2002)[ hep-th/0204187];
 Choudhury, D, D.~Ghoshal, D.~P.~Jatkar and S.~Panda, arXiv:hep-th/0204204;
G Shiu,  and I. Wasserman, 
Phys.Lett. B541 (2002) 6.  G Shiu, S.-H. Henry Tye, I. Wasserman, Phys. Rev. D67 (2003) 083517.  
 Padmanabhan, T., T.~Roy Choudhury,
  Phys.Rev. \emph{D66}, (2002) 081301. arXiv:hep-th/0205055; Padmanabhan, T., Phys. Rev. D{\bf 66}, 021301(2002)[hep-th/0204150].
 Kofman, L and A. Linde, arXiv: hep-th/0205121;
 Sami, M., arXiv:hep-th/0205146;
 Sami, M, P.~Chingangbam and T.~Qureshi, arXiv:hep-th/0205179;
  J. Hwang, H. Noh, Phys.Rev. D66 (2002) 084009; Akira Ishida, Shozo Uehara, Phys.Lett. B544 (2002) 353-356;
N. Moeller, B. Zwiebach, JHEP 0210 (2002) 034;
  Piao, Y.S, R.~G.~Cai, X.~m.~Zhang and Y.~Z.~Zhang, arXiv:hep-ph/0207143;
 Li,  X.Z, D.~j.~Liu and J.~g.~Hao, arXiv:hep-th/0207146;
 Cline, J.M, H.~Firouzjahi and P.~Martineau, arXiv:hep-th/0207156;Yun-Song Piao, Qing-Guo Huang, Xinmin Zhang and Yuan-Zhong Zhang,
 hep-ph/0212219; Zong-Kuan Guo, Yun-Song Piao, Rong-Gen Cai, Yuan-Zhong Zhang,  hep-ph/0304236; Yun-Song Piao, Rong-Gen Cai, Xinmin Zhang, Yuan-Zhong Zhang, hep-ph/0207143;
G. Felder, Lev Kofman and A. Starobinsky, JHEP 0209 (2002) 026[hep-th/0208019].
Wang, B, E.~Abdalla and R.~K.~Su, arXiv:hep-th/0208023;
 Mukohyama, S,arXiv:hep-th/0208094;
   Jian-gang Hao, Xin-zhou Li,  hep-th/0209041;  G.A. Diamandis, B.C. Georgalas , N.E. Mavromatos, E. Papantonopoulos,  hep-th/0203241;
 G.A. Diamandis, B.C. Georgalas , N.E. Mavromatos,
 E. Papantonopoulos, I. Pappa,  hep-th/0107124; M. C. Bento, O. Bertolami and A.A. Sen, hep-th/020812;  M.C. Bento, O. Bertolami., A.A. Sen, Phys.Rev.D67:023504,2003;
    gr-qc/0204046; M.C. Bento, O. Bertolami., Phys.Rev.D65:063513,2002; astro-ph/0111273; 
Jian-gang Hao, Xin-zhou Li, Phys.Rev. D66 (2002) 087301; 
Chanju Kim , Hang Bae Kim and Yoonbai Kim,
hep-th/0210101;  Chanju Kim, Yoonbai Kim, O-Kab Kwon, Chong Oh Lee, hep-th/0305092
; Haewon Lee, W. S. l'Yi, hep-th/0210221; J.S.Bagla,  H.K.Jassal, T.Padmanabhan, astro-ph/0212198;
 M. Sami, Pravabati Chingangbam and Tabish Qureshi, hep-th/0301140; 
G.W. Gibbons, arXiv: hep-th/0301117; 
Chanju Kim, Hang Bae Kim, Yoonbai Kim and O-Kab Kwon, hep-th/0301142; F. Leblond and  A. W. Peet, hep-th/0303035;
F. Leblond, A. W. Peet, hep-th/0305059;  Xin-zhou Li, Dao-jun Liu, Jian-gang Hao, hep-th/0207146;  Xin-zhou Li, Jian-gang Hao, Dao-jun Liu, Chin. Phys. Lett. 19 (2002) 1584;
Tomohiro Matsuda, hep-ph/0302035; Tomohiro Matsuda, hep-ph/0302078;
 A. Das and A. DeBenedictis, gr-qc/0304017;
 Mahbub Majumdar, Anne-Christine Davis, hep-th/0304226; D. Choudhury, D. Ghoshal,  Dileep P. Jatkar (1), S. Panda, hep-th/0305104.
\bibitem{tachyon1} A. Mazumdar, S. Panda, A. Perez-Lorenzana,
Nucl. Phys. B 614 (2001) 101.
\bibitem{babak} S. V. Babak, L. P. Grishchuk, gr-qc/0209006; A. A. Logunov, The Theory of Gravity, gr-qc/0210005 and references therein;  T. Damour, Ian I. Kogan and  Antonios Papazoglou, hep-th/0206044; T. Damour and Ian I. Kogan, hep-th/0206042; M. Sami, hep-th/0210258
; S.S. Gershtein, A.A. Logunov, M.A. Mestvirishvili and N.P. Tkachenko, astro-ph/0305125.
\bibitem{sccc} S. M. Carroll, Living Rev. Rel.{\bf 4}, 1(2001)[astro-ph/0004075].
\bibitem{ratra}  P. J. E. Peebles and Bharat Ratra, Rev.Mod.Phys. 75 (2003) 599-606[astro-ph/0207347].

\bibitem{tpcc} T. Padmanabhan, (2002) {\it Cosmological constant - the Weight of the Vacuum}, to appear
in Phys.Repts [hep-th/0212290].
\bibitem{ph2} V. Sahni and A. A. Starobinsky, Int. J. Mod. Phys. D9,373(2002).
\bibitem{ph1} R.R. Caldwell, Phys.Lett. B54523-29(2002).
\bibitem{ph3} L. Parker and A. Raval, Phys. Rev. D60, 063512(1999).
\bibitem{ph4} T. Chiba, T. Okabe and M. Yamaguchi, Phys. Rev. D62, 023511(2000).
\bibitem{ph5} B. Boisseau, G. Esposito-Farese, D. Polarski and A. A. Starobinsky, Phys. Rev. Lett.85, 2236,
(2000).
\bibitem{ph6} A. E. Schulz, Martin White, Phys.Rev. D64 (2001) 043514.
\bibitem{ph7} V. Faraoni, Int. J. Mod. Phys. D64, 043514 (2002). 
\bibitem{ph8} I. Maor, R. Brustein, J. Mcmahon and P. J. Steinhardt, Phys. Rev. D65 123003(2002).
\bibitem{shatanov} Yu. Shtanov and V. Sahni, astro-ph/0202346.
\bibitem{ph9} V. K. Onemli and R. P. Woodard, Class. Quant. Grav. 19, 4607(2002).
\bibitem{ph10} D. F. Torres, Phys. Rev. D66, 043522 (2002).
\bibitem{hann} S. Hannestad, E. Mortsell, Phys. Rev. D66, 063508 (2002).
\bibitem{t1} A. Melchiorri, L. Mersini, C. J. Odman, M. Trodden,
astro-ph/0211522.
\bibitem{t2} S. M. Carroll, M. Hoffman and  M. Trodden,  astro-ph/0301273.
\bibitem{model} R. Mainini, A.V. Maccio', S.A. Bonometto, A.Klypin, astro-ph/0303303.
\bibitem{ph12} P. H. Frampton, hep-th/0302007.
\bibitem{ph13}Jian-gang Hao and Xin-zhou Li, hep-th/0302100.
\bibitem{ph14} R. R Caldwell, M. Kamionkowski and N. N. Weinberg, 
astro-ph/0302506
\bibitem{ph15} G. W. Gibbons, Phantom matter and the cosmological constant, hep-th/0302199.
\bibitem{ph16} Jian-gang Hao, Xin-zhou Li, hep-th/0303093.
\bibitem{ph17} Shin'ichi Nojiri and  Sergei D. Odintsov, hep-th/0303117; Shin'ichi Nojiri and Sergei D. Odintsov ,hep-th/0304131.
\bibitem{ph18}  Alexander Feinstein and Sanjay Jhingan, hep-th/0304069.
\bibitem{ph19} J. A. S. Lima, J. V. Cunha and  S. Alcaniz, astro-ph/0303388.
\bibitem{ph20} A. Yurov, astro-ph/0305019.
\bibitem{ph21} B. Mcinnes, hep-th/0305107; JHEP 0208 (2002) 029; astro-ph/0210321; JHEP 0212 (2002) 053.
\bibitem{ph22} Jian-gang Hao, Xin-zhou Li,   hep-th/0305207.
 \bibitem{h} F. Hoyle, Mon. Not. R. Astr. Soc. {\bf 108}, 372 (1948); {\bf 109}, 365 (1949).
\bibitem{hn} F. Hoyle and J. V. Narlikar, Proc. Roy. Soc. {\bf A282}, 191 (1964); Mon. Not. R. Astr. Soc. {\bf 155}, 305 (1972); 
{\bf 155}, 323 (1972).
\bibitem{pn} Narlikar, J. V. and Padmanabhan, T., Phys. Rev.D{\bf 32}, 1928(1985).
\bibitem{nb} Hoyle, F., Burbidge, G. and Narlikar, J., V., {\it A Different Approach to Cosmology}, Cambridge Univ. Press(2000).
\bibitem{vishwa} J.V. Narlikar, R.G. Vishwakarma and G. Burbidge, Apj.{\bf 585}, (203)1 [astro-ph/0205064].
\bibitem{vishwa1} R.G. Vishwakarma, Mon.Not.Roy.Astron.Soc. 331 (2002) 776-784. 
\bibitem{matt} M. Visser, {\it Lorentzian Worholes from Einstein to Hawking}, AIP Press (1995).
\bibitem{matt2} M. Visser, S. Kar and N. Dadhich, {\it To appear in PRL},  gr-qc/0301003.
\bibitem{sami} M. Sami, T. Padmanabhan, hep-th/0212317.
 
\bibitem{perl} S. Perlmutter et al., Ap. J. 517, 565 (1999). 
 
\bibitem{riess} A. G. Riess et al., Ap. J. 560, 49 (2001).

\bibitem{blakeslee} J. P. Blakeslee et al.,  astro-ph/0302402. 

\bibitem{wmap} D. N. Spergel et al., astro-ph/0302209.

\bibitem{vishwa2} R. G. Vishwakarma, astro-ph/0302357.



\end{thebibliography}
\end{document}